\documentclass[10pt,letterpaper]{article}
\usepackage{mathrsfs}
\usepackage{graphicx}
\usepackage{multirow}
\usepackage{mathtools}
\usepackage[justification=centering]{caption}
\usepackage{tabularx}
\usepackage{amsmath}
\usepackage{breqn}
\usepackage{booktabs}
\usepackage{hyperref}
\usepackage[lined,commentsnumbered]{algorithm2e}

\begin{document}%
%%%%%%%%%%%%%%%%%
{\Large
\textbf\newline{Multi-Kernel LS-SVM Based Integration Bio-Clinical Data Analysis and Application to Ovarian Cancer} 
}
\newline

Jaya Thomas\textsuperscript{1,2},
Lee Sael\textsuperscript{1,2*}
\\
\bigskip
\textbf{1} Department of Computer Science, SUNY Korea, Incheon, Republic of Korea
\\
\textbf{2} Department of Computer Science, Stony Brook University, Stony Brook, NY, USA
\\
%\textbf{3} Affiliation Dept/Program/Center, Institution Name, City, State, Country
%\\
\bigskip

\section*{Abstract}
The medical research facilitates to acquire a diverse type of data from the same individual for a particular cancer. Recent studies show that utilizing such diverse data results in more accurate predictions. The major challenge faced is how to utilize such diverse data sets in an effective way.
In this paper, we introduce a multiple kernel based pipeline for integrative analysis of high-throughput molecular data (somatic mutation, copy number alteration, DNA methylation and mRNA) and clinical data. We apply the pipeline on Ovarian cancer data from TCGA. After multiple kernels have been generated from the weighted sum of individual kernels, it is used to stratify patients and predict clinical outcomes.
We examine the survival time, vital status, and neoplasm cancer status of each subtype to verify how well they cluster. We have also examined the power of molecular and clinical data in predicting dichotomized overall survival data and to classify the tumor grade for the cancer samples. It was observed that the integration of various data types yields higher log-rank statistics value. We were also able to predict clinical status with higher accuracy as compared to using individual data types.

\section{Introduction}

Cancer is a disease with extreme complexity which alters the function of combinations of genes. It is believed to be an outcome of accumulated genetic changes \cite{Aunoble00}. Among various types of cancer, ovarian cancer is the fifth most common cancers diagnosed in females \cite{Ries07} with overall five-year survival rate only around 44$\%$\cite{OvCS}. The Cancer Genome Atlas (TCGA)\cite{TCGA} reports diverse genomic information with paired clinical information for more than 500 cases of ovarian serous cystadenocarcinoma.
The genomic information includes copy number alteration (CNA), somatic mutation, gene expression, and DNA methylation. Understanding the genetic changes in cancer patients through this rich information allows for better diagnostics and treatment of cancer, including ovarian cancer.

Integrative analysis of multiple perspectives of a patient helps in both patient stratification and clinical outcome prediction. Patient stratification and clinical outcome predictions both help the researchers in understanding and exploring the genomic characteristics in a relationship with their current phenotypes and thus to recognizing opportunities for clinical improvement. In the case of cancer data analysis, including ovarian cancer data, an improved stratification and clinical prediction can be achieved by integrative analysis of the multiple bio-clinical data. However, due to the complex relationship between the multiple data types, the integrative analysis is still a challenging task.

There are several works related to clinical outcome predictions. Wang \emph{et al.} \cite{Wang05} have used gene expression data to predict distant metastasis of lymph-node-negative primary breast cancer. They identified a 76-gene signature consisting of 60 genes for patients positive for estrogen receptors (ER) and 16 genes for ER-negative patients.
Teschendorff \emph{et al.} \cite{Teschendorff06} proposed a gene expression classifier for ER positive breast cancer.
Zhang \emph{et al.} \cite{Zhang09} used copy number alterations in combination with gene expression to identify the genomic loci and their mapped genes, having a high correlation with distant metastasis capability of human breast cancer.
Deneberg \emph{et al.} \cite{Deneberg10} used gene specific and global methylation patterns predict outcome in patients with acute myeloid leukemia. They also concluded in their work that global and gene specific methylation patterns are independently associated with the clinical outcome in AML patients.
Nair \emph{et al.} \cite{Nair12} reported a comprehensive review on the clinical outcome prediction by the miRNA expression for numerous types of cancer. These approaches only integrated a smaller number of data types and failed to integrate with other levels of genomic data.

On the other hand, for the patient stratification, biomarkers, genetic profiles, research data along with clinical information are used to find a subgroup of the patients thereby making easier to  detect and interpret relationships as well as predict outcomes in a specific subgroup.
Kim \emph{et al.}\cite{Kim2014} considers somatic mutation profile and exploited k-means clustering to identify the tumor subtypes.
The sparsity of the mutation data was handled by applying Jaccard and Euclidean distance measures.
Further, the Cox proportional hazards regression model was used to find the similarity between the derived subtypes and the patient survival time.
In their recent work \cite{Kim2015c}, a compressed somatic mutation profile was suggested for fast comparison.
The profile utilized Gene-Ontology and non-negative matrix factorization for condensing the mutation profile.
To verify their work, stratification was performed on various cancer types.
Hofree \emph{et al.} \cite{Hofree13} has used genome-scale somatic mutation profiles in combination with a gene interaction network to carry out subgrouping of patients.
Recently, Wang \emph{et al.}\cite{Wang14} proposed a modified consensus clustering to carry out patient stratification for breast cancer patients.
The approach considered both numerical and categorical data for mRNA and miRNA data set.

Analysis of one or few data types may not be sufficient for accurate predict or stratification.
Thus, efforts to integrate the molecular data were carried out.
Thomas \emph{et al.}\cite{Thomas2015} work presents two general class of heterogeneous data integration, i.e., Multiple Kernel learning and Bayesian network, are detailed and discussed in the bioinformatics domain.
Also, many problem-specific integrative approaches have been proposed to associate the molecular data with the clinical outcome.
These include a software package implemented in R \cite{Louhimo11} to show the effect of DNA methylation and copy number alterations in gene expression of several known oncogenes for two cancer type glioblastoma multiforme and ovarian.
Kim \emph{et al.} \cite{Kim12} proposed a graph based integrated framework using CNA, methylation, miRNA, and gene expression data to carry out a molecular based classification of clinical outcomes.
In this approach, a single graph was constructed by determining the optimum linear combination coefficient from the multiple graphs obtained at different genomic level.
Sohn \emph{et al.} \cite{Sohn13} modeled the influence of multi-layered genomic features on gene expression traits by modeling an integrative statistical framework based on a sparse regression. The results showed that using CNA, miRNA, and methylation on gene expression in the predictive power for gene expression level is improved over a single data type based analysis.
Schafer \emph{et al.} (\cite{Schafer09} approach integrated copy number and gene expression by a modified correlation coefficient and an explorative Wilcoxon test to find DNA regions of abnormalities. The recent work also includes model based prediction of clinical outcomes.
Mankoo \emph{et al.} \cite{Mankoo11} have applied multivariate Cox Lasso model and median time-to-event prediction algorithm on data set integrated from the four genomic data types (CNA, methylation, miRNA, and gene expression data).
Yuan \emph{et al.} \cite{Yuan14} evaluated the predictive power of patient survival and binary clinical outcome using clinical data in combination with one molecular data: somatic copy number alteration, DNA methylation, and mRNA, miRNA and protein expression. They showed a slight improvement in some cases when clinical information was combined with one of the molecular. Although this paper showed the predictive power of clinical data in combination with a molecular data, all available molecular data was not used integratively.

%Many visual analysis, including StratomeX \cite{Lex12} and iGPSe \cite{Ding14}, are also proposed. StratomeX\cite{Lex12} stratifies patients across multiple genomic data types such as gene expression, DNA methylation, or copy number data. These subtypes are further explored to determine functional and clinical implications. iGPSe\cite{Ding14} is a web based visual analytical system used for patient stratification and exploration of disease subtypes in heterogeneous genomic data. In this system, the comparison is done using survival analysis with the help of unsupervised clustering with graph and parallel sets visualization. The network models are also used for patient stratification.

\begin{figure*}[!tpb]%figure0
\includegraphics[height=8cm, width=10cm ]{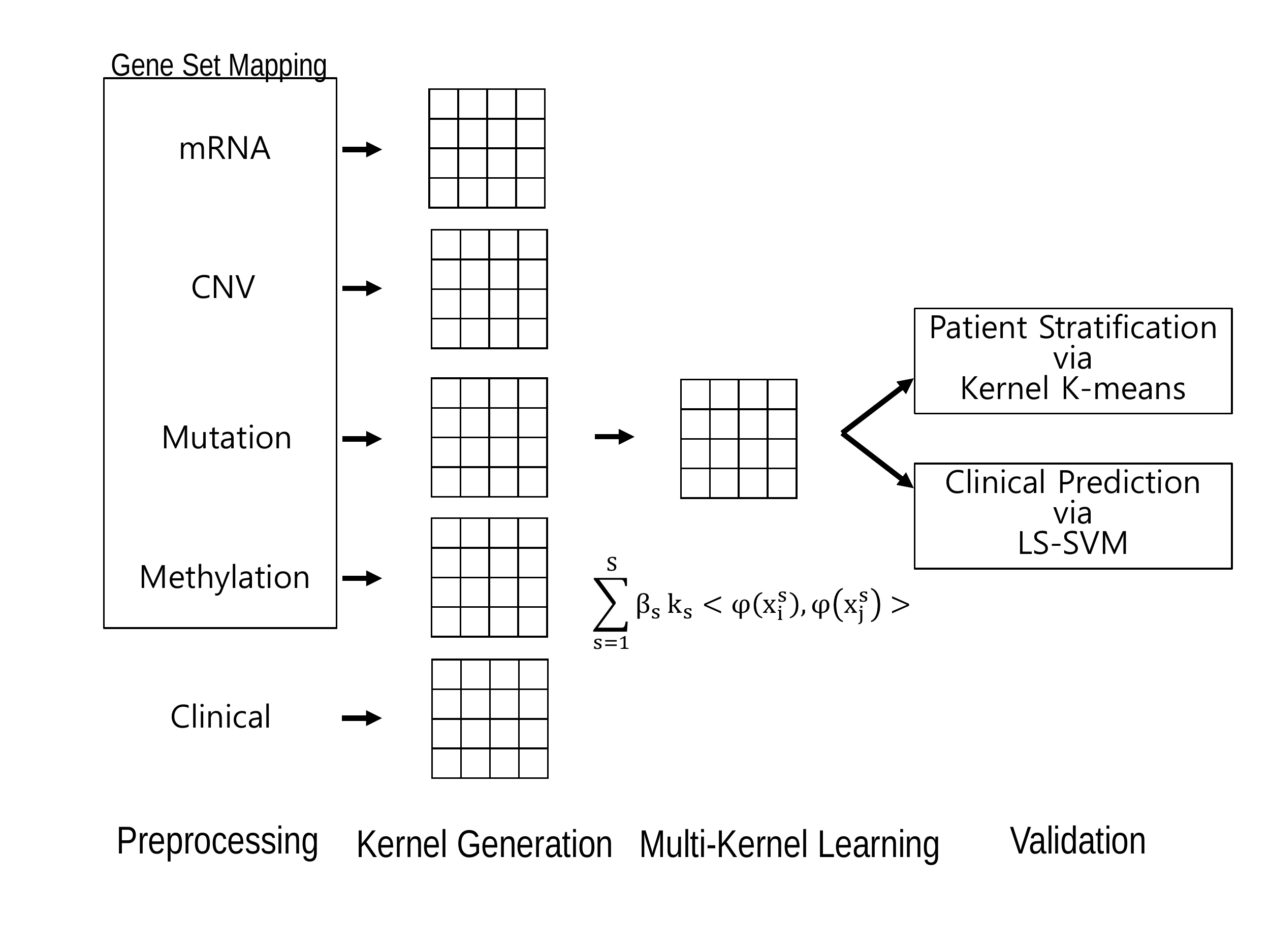}
\centering
\caption{ Multi kernel learning based  integrative pipelined model}
\label{fig:00}
\end{figure*}

An integrative analysis method that can cover heterogeneity of data types in molecular data and clinical data can beneficial in predicting the prognostics of patients via stratifying the patients in the different risk groups. Multiple kernel learning is well known for addressing various data heterogeneity. Moreover, Kernel methods, including multiple kernels, are well-suited for handling non-linearity of high dimensional data by mapping data to feature space \cite{Bucak14}.

In this paper, we make the following contributions:
\begin{itemize}
\item{\textbf{Combines clinical data with multiple molecular data}. We examine how adding more molecular information increases the prediction performance in stratifying ovarian cancer patients, and predicting tumor grade and patient survival time.}

\item{\textbf{Propose a multiple kernel based pipeline model (Fig. \ref{fig:00}) to integrate multiple heterogeneous data types.}  The proposed model allows to analyze heterogeneous data  i.e., combines data with diverse background distributions, relations, dimensions, and formats to enhance the statistical significance and thus, obtain more refined information. }

\item{\textbf{Propose the data pre-processing using patient-centered gene set analysis.} It allows to handle the large heterogeneous tumor data by grouping them into much smaller set of pathways and biologic processes.}

\end{itemize}

\section{Material and methods}

\subsection{Datasets and Raw Mutation Scores}
\par
%\kant[1-2]

%\begin{table}[h]
%\caption{Numbers of samples and features of data types for OV cancer.}
%\centering
%\setlength{\tabcolsep}{.2\tabcolsep}
%\resizebox{\columnwidth}{!}{
%\begin{tabular}{|l|l|l|l|l|}
%\hline
%Data Type
%& Platform
%& \multicolumn{1}{|p{2cm}|}{\centering $\#$Genes %altered in \\ 586 patient}
%& \multicolumn{1}{|p{2cm}|}{\centering $\#$Genes altered in \\ 312 patient}
%& \multicolumn{1}{|p{2cm}|}{\centering Union of considered genes}\\ \hline
%Methylation&Illumina Human Methylation&22580&13772&13772\\
%CNA &Agilent 1M&20175&16383&16070\\
%mRNA expression&AgilentG4502A&20531&18361&16070\\
%Mutation&WUSM (Mutation calling)&15917&9039&9039\\
%\hline
%\end{tabular}
%}
%\label{Tab:01}
%\end{table}

Data are initially selected and downloaded 312 samples that contained all four genomic data types, i.e., copy number alternation, methylation, mRNA expression and the mutation information, from TCGA data portal \cite{tcgalink} via TCGA assembler \cite{Zhu14} and TCGA Firehose \cite{linkgdac}.
The summary of the genomic data types and number of associated genes for each data type in the 312 samples are shown in Table~\ref{Tab:01}.
Clinical information of the 312 samples is also downloaded from TCGA. The clinical data includes the survival time (days to death), age, tumor stage, tumor grade, vital status and neoplasm cancer status.

\begin{table*}[h]
\caption{Numbers of samples and features of data types for OV cancer.}
\centering
\begin{tabular}{|l|l|l|l|l|}
\hline
Data Type
& Platform
& \multicolumn{1}{|p{2cm}|}{\centering $\#$Genes altered in \\ 312 patient}
& \multicolumn{1}{|p{2cm}|}{\centering Union of considered genes}\\ \hline
Methylation&Illumina Human Meth.&13772&13772\\
CNA &Agilent 1M&16383&16070\\
mRNA expression&AgilentG4502A&18361&16070\\
Mutation&WUSM&9039&9039\\
\hline
\end{tabular}
\label{Tab:01}
\end{table*}
%\kant[3-5]

Description of the data types and how each is further processed are provided in the following.
For each DNA methylation sample, the percent signal that is methylated is described as beta value recorded for each sample locus. The beta values are continuous variables that range between 0 and 1 indicating the ratio of the intensity of the methylation \cite{Liu13}. After downloading level three data from TCGA assembler \cite{Zhu14}, the data is pre-processed and determined to be methylated, if they show a percentage of methylation (beta) greater than a certain threshold ( 0.3 for patient data) or unmethylated if the value falls below the threshold as discussed by
Warden et al. \cite{Warden13}.
Thus, the new data matrix constructed for the methylation has a score of 1, if a gene is methylated else set to 0.
The level three data for copy number alternation (CNA) was obtained from the  GISTIC \cite{Beroukhim07} analysis.
GISTIC identifies genomic regions that are significantly gained or lost across a set of tumors. It contains data about the significant regions of amplification and deletion as well as which samples are amplified or deleted in each of these regions.
The matrix element with a value of 0 indicates no amplification or deletion above the threshold. Amplifications are positive numbers: 1 denotes amplification above the amplification threshold; 2 denotes amplifications larger to the arm level amplifications observed for the sample.
Deletions are represented by negative table values: -1 represents deletion beyond the threshold; -2 represents deletions greater than the minimum arm-level deletion observed for the sample. The data matrix generated from CNA data puts 1 if the gene is amplified or deleted and 0 if otherwise.
The dataset downloaded for level 3 mRNA from TCGA Firehose \cite{linkgdac} contains $log2$ ratio for the gene expression.The $log2$ ratio ranges from 0 to 16, representing relative gene expression levels.
The level 2 somatic mutation data download from TCGA is already in the required matrix format with the entries showing either 0 or 1 indicating the presence or absence of a mutation in the gene.

\subsection{Gene Sets and Adjusted Mutation Scores}

We group the genes based on involvement in the same pathway or having the similar molecular signature, thus some of the genes that do not fall in these categories were filtered out. The total number of genes considered in this study are summarized in the last column of Table \ref{Tab:01}.

Considering gene set takes into account the fact that genes do not act in isolation, but they interact with other genes through the complex system. Also, cancer occurs not in a single gene, but rather, a group of genes that interact amongst each other in the complex biological network  \cite{Kim2014, Kim2015c}. Moreover, the biological significance can be better analyzed by considering the interaction between neighboring genes. For the different data sources, this measure helps to construct a patient to geneset matrix containing the genomic information. We have considered the functional group information of genes initially downloaded from the Molecular Signatures Database (MSigDb) \cite{Subramanian05} and recreated to remove redundancy. In the MSigDB, we select the group information based on pathway (C2: 4722 gene sets) and based on motif (C3: 836 gene sets). MSigDB contains gene sets generated from KEGG \cite{Kanehisa00}, Canonical Pathway \cite{Liberzon11}, BIOCARTA \cite{Nishimura01} and  REACTOME \cite{Croft14}. The motif gene set contained in MSigDB are miRNA targets (MIR) and transcription factor target (TFT).

We recreate the gene sets to generate unified gene groups with small overlaps while maximizing the number of genes covered. We filter out the gene sets with more than 85$\%$ overlap as described in Algorithm \ref{alg:geneset}. After filtering, 2099 gene sets remained and the gene sets cover 16070 genes out of the initial 16095 genes.

\begin{algorithm}[H]
% \SetAlgoLined
 \KwData{Gene sets in C2 and C3}
 \KwResult{Selected gene set in F}
 \BlankLine
 S = \{all sets in C2 and C3 ordered by number of genes in the gene set, smallest to largest\} \\
 F = \{\}  // empty set \\
 \ForEach {s1 in S} {
    SimSet = 0 \\
	 \ForEach {s2 in S - \{s1\}} {
        dist = (s1 $\cap$ s2) / (s1 $\cup$ s2)       // Jaccard similarity \\
		\If {dist > 0.85  } { SimSet++; }
	 }
    \If {SimSet == 0 } { put s1 to F }
  delete s1 from S
 }
 \caption{Gene set selection.}
 \label{alg:geneset}
\end{algorithm}

%Gene set score is assigned based on the number of genes altered in the gene sets in each patient. The higher value for the mutation score for a gene set indicates the severity of the mutation in the relevant pathway or motif. These mutation score values are normalized to range between 0 and 1.
Generated patient-to-gene set matrix contains gene sets as a new feature vector, where each entry is an aggregating value of the altered genes in the gene set.
%To construct the matrix, once genes are grouped the mutation scores are adjusted to reflect the proportion of mutated genes in a gene set.

%\pagebreak
\subsection{Kernel Matrix Representing Molecular Information}

The patient-to-gene set matrixes of the four data sources are used to create kernels using kernel functions. A feature function, $\phi(\mathbf{x})$, maps the original data feature $\mathbf{x}$ in the input space to a high-dimensional feature space. A Kernel function is a function that corresponds to the inner product in a expanded feature space: $k(\mathbf{x}_{i},\mathbf{x}_{j}) = <\phi(\mathbf{x}_i) \cdot \phi(\mathbf{x}_j)> $. A kernel matrix is formed by computing kernel functions between all pairs of data. Thus, the size of a kernel matrix is independent of the number of features and is solely dependent on the number of data. In practice, an explicit definition of feature function, $\phi(\mathbf{x})$, is not needed since they are tightly integrated into the definition of the kernel functions.

The kernel functions we used are linear and radial basis function (RBF).
Details of linear and RBF kernels are as follows:
Let i$^{th}$ and j$^{th}$ sample data be represented as vectors of adjusted mutation scores of each gene sets: $\mathbf{x}_{i}$ and $\mathbf{x}_{j}$.
A linear kernel of two samples is a dot product of their original feature vectors, $\mathbf{x}_{i}$ and $\mathbf{x}_{j}$:
\begin{equation*}
k_{linear}(\mathbf{x}_{i},\mathbf{x}_{j})=<\mathbf{x}_{i} \cdot \mathbf{x}_{j}>.
\label{eq1}
\end{equation*}
A RBF kernel of two samples vectors $\mathbf{x}_{i}$ and $\mathbf{x}_{j}$ is defined as follows:
\begin{equation*}
k_{RBF}(\mathbf{x}_{i},\mathbf{x}_{j})=exp(-||\mathbf{x}_{i}-\mathbf{x}_{j}||^{2}/2\sigma^{2}),
\label{eq2}
\end{equation*}
where $||\mathbf{x}_{i}-\mathbf{x}_{j}||^2$ is the squared Euclidean distance between the two original feature vectors and parameter $\sigma$ controls the flexibility of the kernel. With smaller value for the parameter $\sigma$, the kernel matrix becomes closer to identity matrix while risking overfitting. On the other hand, larger values of parameter gradually reduce the kernel to a constant function, making it impossible to learn any non-trivial classifier \cite{Cristianini04}.
In our experiment, we use a separate validation data sets consisting of 25$\%$ of total samples to determine the parameters of the kernels, such as the value of $\sigma$ in RBF kernels. The choice of the kernel for different data sources is decided based on the existing study in the literature.

We explored the use of commonly used kernels including linear, sigmoid, polynomial and the radial basis function. We chose the kernel function that showed the best performance for each data type.
%In addition, we also considered the available literature knowledge for selection of the kernel function.
For mRNA, we select RBF kernel as in \cite{Gomes2010}, the authors report in their work that the use of RBF kernel proved to be more effective as compared to linear, polynomial or other kernels. For a different combination of geneset, an accuracy of 92.59 $\%$ was observed. The methylation data analysis \cite{Zhang15} shows that the use of SVM classifier with RBF kernel outperforms the k-nearest neighbor classifier (k-NN) and a naive Bayes classifier with an accuracy of 91.3 $\%$.  In \cite{Thomas14}, the performance of RBF kernel is compared to the clinical kernel for the clinical data source. Out of the five case study carried out, it was observed that the RBF kernel outperforms clinical in three with an average accuracy of 78.59 $\%$. In the case of CNV, we apply linear kernel, similar to \cite{Seoane14} which uses linear kernel for CNV data source for classification and attain an accuracy of 61$\%$. For mutation data source, we apply RBF kernel, as in \cite{Pirooznia06} presents a detailed  comparison result for linear, polynomial and RBF kernel functions for breast cancer mutation data. It is reported that the use of RBF kernel for classification achieved a higher accuracy as compared to other kernels. For BRCA1-BRCA2 dataset, RBF attained an accuracy of 100$\%$ as compared to 93.3$\%$ (linear) and 86.6 $\%$ (polynomial).

\subsection{Multiple Kernel Learning for Cancer Classification}

The kernel matrix constructed from each data types is further integrated to form a single kernel matrix using a multiple kernel learning approach. Several methods are suggested for integrating the kernels \cite{Gonen2011}. We take a two-step approach that first combines the kernels in a weighted linear fashion and then perform learning on the combined kernel. The kernel combination is defined as follows:

\begin{equation*}\label{eq:1}
K_{\beta}(\mathbf{x}_{i},\mathbf{x}_{j})= \sum\limits_{s=1}^S \beta_{s}k_{s}(\phi(\mathbf{x}_{i}^{s}),\phi( \mathbf{x}_{j}^{s}))
\end{equation*}
\begin{equation*}
\text{subjected to  } \beta_{s} \geq 0 \text{ and } \sum\limits_{s=1}^S \beta_{s}=1,
\end{equation*}
where $S$ is the number of kernels, $\mathbf{x}_{i}^{s}$ is the original feature vector of kernel $s$ of sample $i$, and $\beta_n$ is the kernel coefficient of kernel $s$.

To obtain optimal weights for kernel combination, we take the optimization approach suggested by Zien \emph{et al.} \cite{Zien07}. In their approach, the kernel coefficient is determined by the efficacy of each of the kernel matrix containing sets learned by Least Square Support Vector Machine (LS-SVM). LS-SVMs are closely related to regularization networks and Gaussian processes but additionally emphasize and exploit primal-dual interpretations from the optimization theory \cite{Suykens99}. The primal form of a LS-SVM is optimized by the following minimization problem:
\begin{equation*}
\min\limits_{\mathbf{w},b,err}(\frac{1}{2} \mathbf{w}^{T}\mathbf{w}+\gamma\sum\limits_{i=1}^N err_{i}^{2})
\end{equation*}
\begin{equation*}
\textrm{subjected to   }
y_{i}[\mathbf{w}^{T} \phi(\mathbf{x}_{i})+b]=1-err_{i}^2 \text{    for } i=1,2,\dots,N
\label{eq:LSSVM}
\end{equation*}
where $\mathbf{w}$ is the weight vector we are trying to learn, $err_{s}$ is the error variables that represent the value corresponding to misclassification in case of overlapping distribution, and $\gamma$ is the regularization parameter that tackles data over fitting problem.

The standard multiple kernel learning approach constructs the base kernels for each data type and determine their optimal kernel coefficient by solving Equation \ref{eq:Multi-LSSVM} \cite{Shi12, Yeh12}.
\begin{equation*}
\min\limits_{\beta,\mathbf{w},b,err} (\frac{1}{2} \sum\limits_{s=1}^S \beta \mathbf{w}^{T}\mathbf{w} + \gamma\sum\limits_{i=1}^N err_{i}^{2})
\label{eq:Multi-LSSVM}
\end{equation*}
\begin{equation*}
\begin{split}
\textrm{w.r.t.  } &
w_{k} \in \Re^{D_{k}}, err \in \Re^{N} \\
\end{split}
\end{equation*}
\begin{equation*}
\begin{split}
\textrm{subjected to   }
\begin{cases}
y_{i}(\sum\limits_{s=1}^S \beta_{s}\mathbf{w}_{s}^{T} \phi_{k}(\mathbf{x}_{i})+b)= 1-err_{i}^2,\\
err_{i} \geq 0 \text{    for } i=1,2,\dots,N,\\
\sum\limits_{s=1}^S \beta_{s}=1, \beta_{s}\geq 0 \text{     for  } s=1,2,\dots,S. \\
\end{cases}
\end{split}
\end{equation*}

The derived dual for the problem in Equation \ref{eq:Multi-LSSVM} \cite{Bach2004}is given as

\begin{equation*}
\min    \delta - \sum\limits_{s=1}^S \alpha_{i}
\label{eq:Dual-LSSVM}
\end{equation*}
\begin{equation*}
\delta \in \Re, \alpha \in \Re^{N}
\end{equation*}
\begin{equation*}
\begin{split}
\textrm{subjected to   }
\begin{cases}
0\leq \alpha \leq 1\gamma, \sum\limits_{i=1}^S \alpha_{i}y_{i}=0 \\
\frac{1}{2} \sum\limits_{i,j=1}^S \alpha_{i}\alpha_{j}y_{i}y_{j}k_{k}(x_{i},x_{j})\leq \delta, \forall=1,\dots,K
\end{cases}
\end{split}
\end{equation*}

The standard multiple kernel learning approach constructs the base kernels for each data type, and determine their optimal kernel coefficient by solving Equation \ref{eq:LSSVM}. Here, the optimization problem is solved using semi-defined linear programming.
The dual for the problem is computed by considering the problem ($D_{k}$), squaring the constraints $\delta$, multiplying the constraints by $\frac{1}{2}$ and performing substitution as $\frac{1}{2} \delta^{2} \mapsto \delta$ leads to dual form of multiple kernel learning Equation \ref{eq:Dual-LSSVM}, here $k_{k}(x_{i},x_{j})=<\phi_{k}(x_{i}),\phi_{k}(xj)>$. This process uses transductive learning setting, where the kernel matrix is learned from data. Initially labeled training data is used to learn the good embedding, which is later applied to unlabeled test data. Considering semi-defined linear programming optimization using SVM enables to handle the optimization of convex cost functions and machine learning concerns, thus provides a powerful method for learning the kernel matrix \cite{Lanckriet04}.

\subsection{Stratification Using Kernel K-means}

Stratification of patients can be done with clustering methods. We use kernel K-means on the generated multiple kernel matrix for stratifying the ovarian cancer to subtypes. The multiple kernel matrix contains the similarity information about pairs of data in the combined feature space. Thus, when we apply the kernel k-means to the multiple kernel matrix, data are clustered so that the clustering error is minimized in the combined feature space. The objective function of kernel k-means is defined as follows:
\begin{equation*}
D({\{\pi_{c}}\}_{c=1}^{k})=\sum\limits_{c=1}^k\sum\limits_{\mathbf{x}_{i}\in \pi_{c}}  ||\phi(\mathbf{x}_{i}) - \mathbf{m}_c||^2,\\
\text{where} \;\; \mathbf{m}_{c}=\frac{\sum\limits_{\mathbf{x}_{i}\in \pi_{c}}\phi(\mathbf{x}_{i})}{|\pi_{c}|},
\end{equation*}
where $\pi_{c}$ denotes the clusters, ${\{\pi_{c}}\}_{c=1}^{k}$ denotes a partitioning of points, $\mathbf{m}_{c}$ denotes the center of cluster $\pi_{c}$, and $|\pi_{c}|$ denote the size of the cluster $\pi_{c}$.
The Euclidean distance between the data point, $\phi(\mathbf{x}_{i})$, and the cluster center, $\mathbf{m}_{c}$, in the feature space is determined as follows \cite{Lanckriet04}:
\begin{equation*}
||\phi(\mathbf{x}_{i}) - \mathbf{m}_c|| = \\
\phi(\mathbf{x}_{i})\phi(\mathbf{x}_{i})
- \frac{\sum\limits_{\mathbf{x}_{j}\in\pi_{c}}\phi(\mathbf{x}_{i})\phi(\mathbf{x}_{j})}{|\pi_{c}|}
+ \frac{\sum\limits_{{\mathbf{x}_{j}, \mathbf{x}_{l}} \in \pi_{c}}\phi(\mathbf{x}_{j})\phi(\mathbf{x}_{l})}{|\pi_{c}|^{2}}
\end{equation*}

Here, the $\phi(x_{i})\phi(x_{j})$ is computed using appropriate selected kernel functions.

\subsection{Clinical Feature Prediction Process}

We also use the learned multiple kernel matrix for predicting clinical outcomes.
For prediction, we again employ LS-SVM classifier.
That is, using the multiple kernel as input, we run the LS-SVM to predict the survival time and the tumor grade of patients.
The performance of the proposed model was evaluated on 312 samples in TCGA ovarian cancer data sets. Each of the samples contains sets of molecular data with matched clinical information.
We split the 312 sample randomly so that $50\%$ of the samples are assigned to the training set, $25\%$ assigned to the validation set to learn the model parameters, and rest are assigned to the testing set to test the performance of the final model.
For evaluations, we calculate the accuracy of survival prediction and area under the curve (AUC) of the receiver operating characteristic (ROC) curve for the tumor grade classification.
The curves were constructed by plotting true positive rate (Sensitivity) in function of the false positive rate (100-Specificity) for different selected threshold for the tumor grade parameter. We selected the threshold range from 0.2 to 0.9. Each point on the ROC curve denotes a sensitivity/specificity pair corresponding to a selected decision threshold. The area under the curve specifies the ability of the test to correctly classify high grade and low grade tumor. The AUC value is computed by a non-parametric method based on constructing trapezoids under the curve as an approximation of area.

%The AUC of ROC curves can also be interpreted as the average sensitivity over the entire range of possible specificities, or the average specificity over the entire range of possible %sensitivities \cite{Eng05}.
% \cite{Freeman08}.

\section{Results}

We report the results of validation and performance of combining the clinical features with the biological features by the multiple-kernel on two important translational bioinformatics tasks: patient stratification and clinical predictions.

\subsection{Patient Stratification via K-means}
We performed the kernel k-means clustering to stratify ovarian cancer patients using the generated kernel matrices.
We compared four data type combinations as input to the k-mean clustering:
the first multiple kernel is constructed from only the molecular data types listed in Table \ref{Tab:01},
the second is constructed from clinical information (i.e., age, stage, grade),
the third is constructed by a non-weighted linear combination of kernels of molecular as well as clinical data,
and the fourth is construed by weighted linear combination of kernels of molecular and clinical data.

To evaluate the clustering result, we performed survival analysis on each cluster, or subgroups, using the Cox proportional hazards regression model in the R survival package \cite{Therneau14} for each of the data type combinations.
Out of 312 patient samples, the clustering was carried out for $75\%$ (231) of the samples and $25\%$ (81) to determine the number of clusters, k.

\begin{figure*}[t]%figure1
\centering
\includegraphics[height=3.5cm,width=12cm]{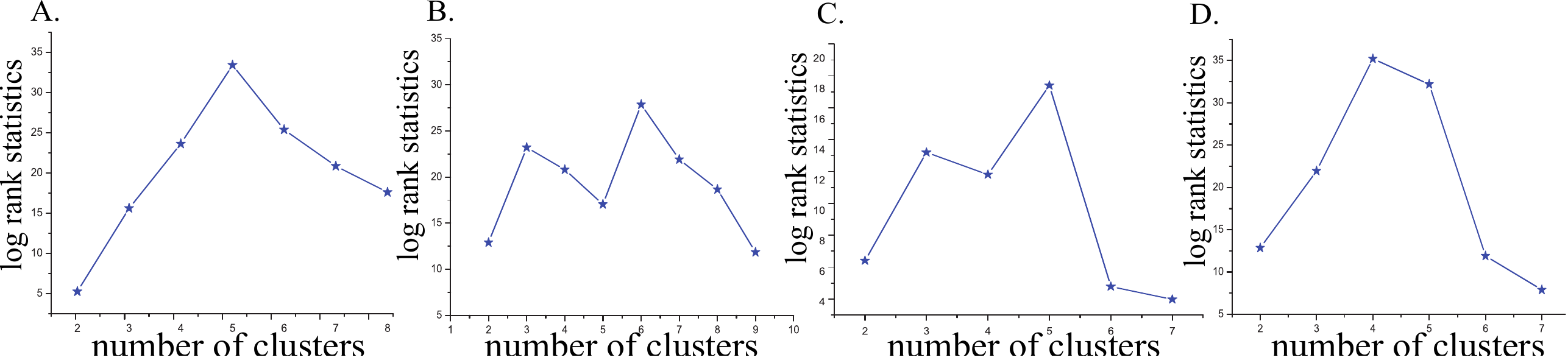}
\caption{Log rank statistic to determine the number of clusters (A) Molecular data  (B) Clinical data (C) Molecular and clinical data with non-weighted linear kernel coefficient (D) Molecular and clinical data with weighted kernel coefficient}
\label{fig:01}
\end{figure*}

The value of k (i.e., the number of clusters) was determined using the log rank statistics.
Figure ~\ref{fig:01} shows the different log rank statistic values obtained for a different number of clusters.
Figure ~\ref{fig:01} (A) shows the plot for integrated molecular data indicating the best value for k is 5.
Figure ~\ref{fig:01} (B) is a graph for determining the k (i.e., 5) value for clinical data.
Similarly, figure ~\ref{fig:01} (C) and figure ~\ref{fig:01} (D) plots results when molecular data is integrated with clinical without and with weighted kernel coefficient, results in best clusters for k=5 and k=6 respectively.
We compared the survival times for these clusters using log-rank statistics and obtained the P-value. The P-value for all the above cases is less than 0.05.
Thus, it shows that there exists a significant separation between the subgroups with respect to survival time.

\begin{figure*}[t]
\includegraphics[height=4cm,width=10cm]{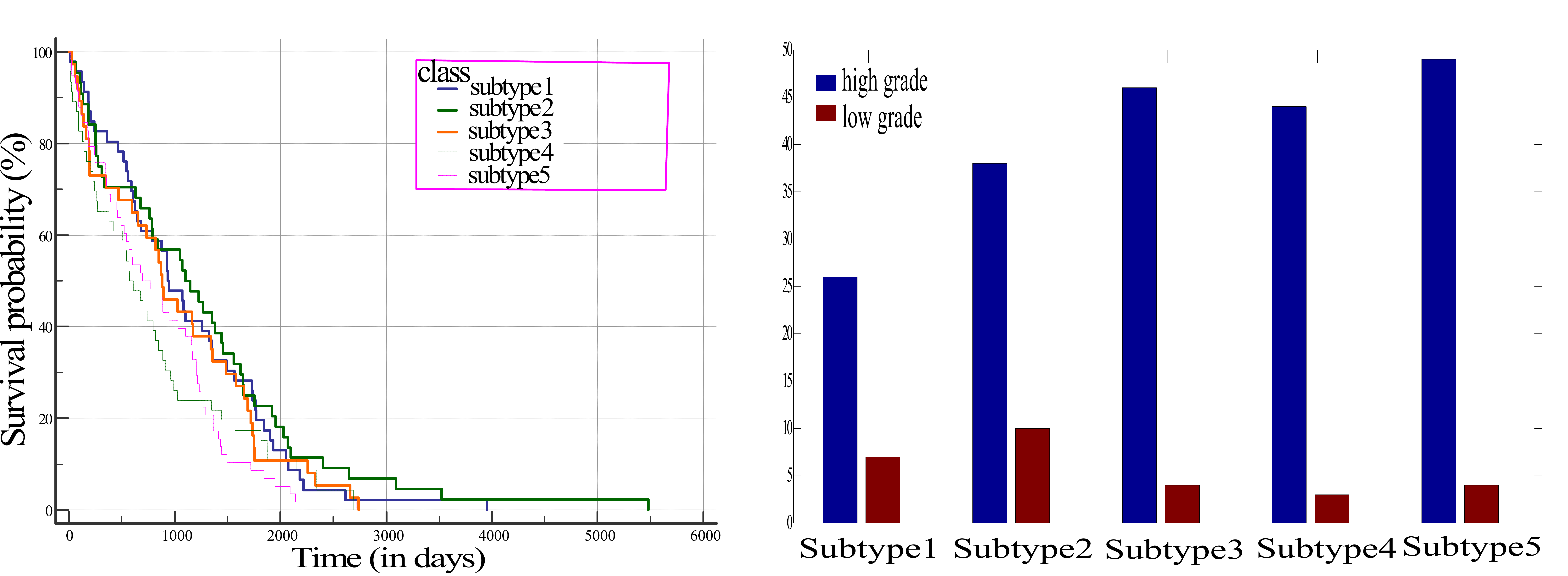}
\centering
\caption{Kernel k-means clustering of the TCGA OV all molecular data reveals (a) five molecular subtypes (clusters) (b) tumor grade for each subtype.}
\label{fig:02}
\end{figure*}

\begin{figure*}[t]
\centering
\includegraphics[height=4cm,width=10cm]{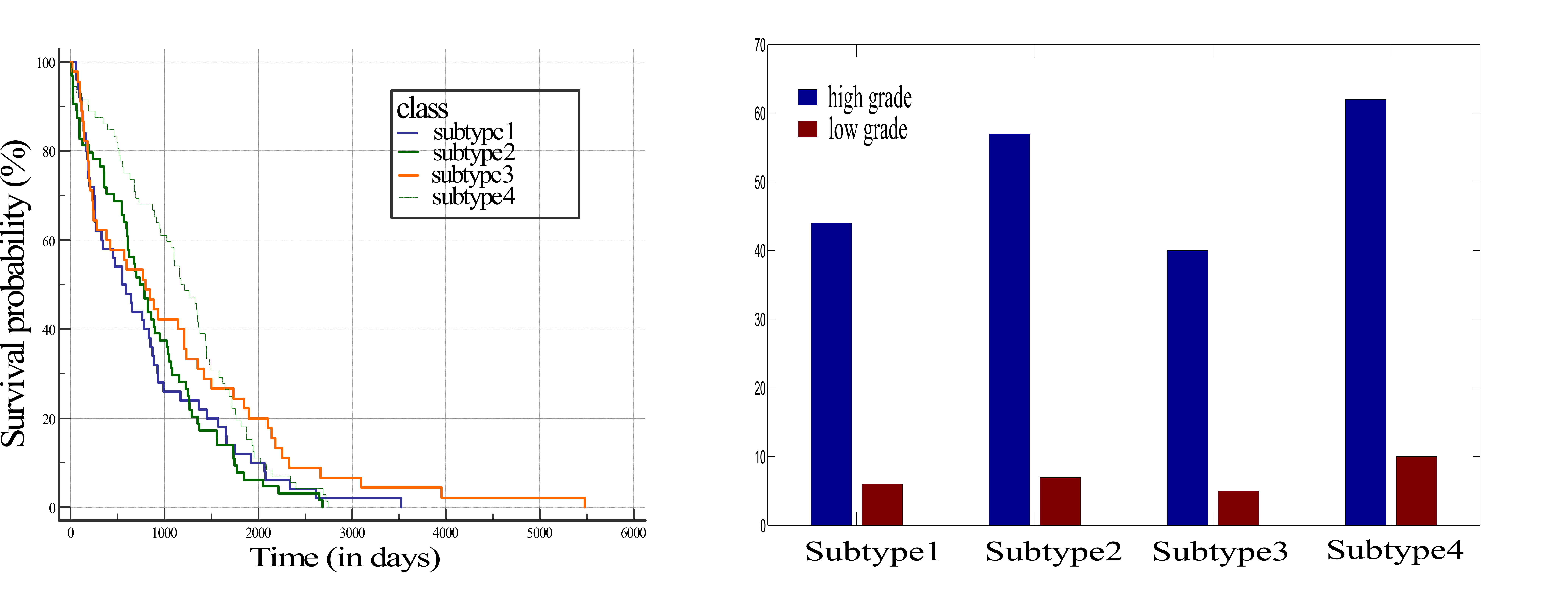}
\caption{Kernel k-means clustering for clinical data}
\label{fig:03}
\end{figure*}

\begin{figure*}[t]
\centering
\includegraphics[height=4cm,width=10cm]{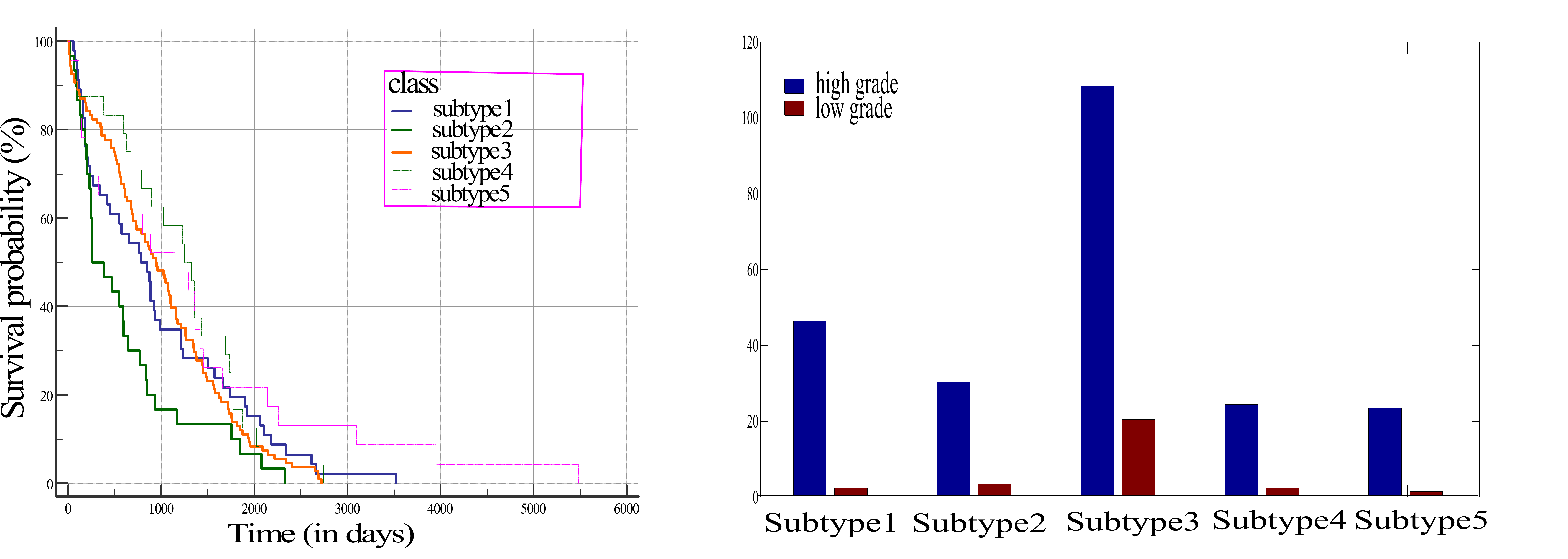}
\caption{Kernel k-means clustering of the TCGA OV all molecular data with clinical data with linear kernel combination reveals (a) five molecular subtypes (clusters) (b) tumor grade for each subtype.}
\label{fig:04}
\end{figure*}

\begin{figure*}[t]
\centering
\includegraphics[,height=4cm,width=10cm]{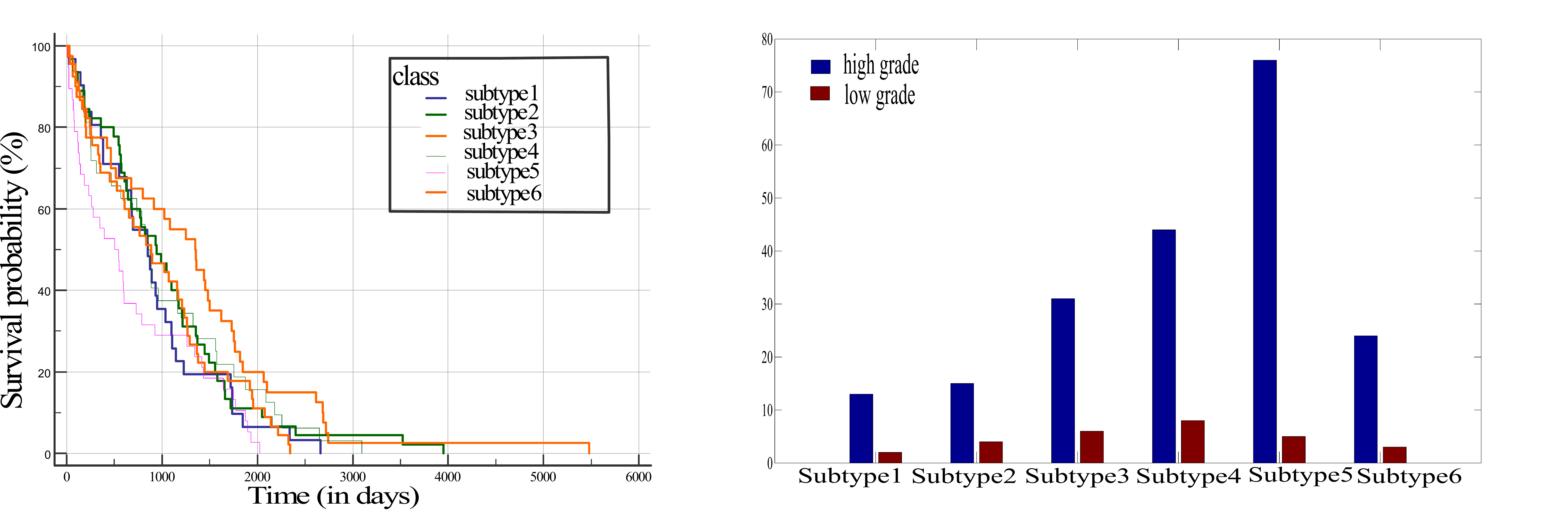}
\caption{Kernel k-means clustering of the TCGA OV all molecular data with clinical data with optimized weights reveals (a) six molecular subtypes (clusters) (b) tumor grade for each subtype.}
\label{fig:05}
\end{figure*}

Setting k=5 in kernel k-means clustering, the p-value of the subtype separation for survival analysis is 0.02 for all molecular data types (figure ~\ref{fig:02} (a)), 0.0079 for clinical data (figure ~\ref{fig:03} (a)), 0.009 for integrated molecular and clinical data with non-weighted kernel coefficient (figure ~\ref{fig:04} (a)), 0.0014 for integrated molecular and clinical data with weighted kernel coefficient (figure ~\ref{fig:05} (a)). It can be observed that the clusters identified by integrating the clinical data are more predictive with log-rank p-value of 1.4 x 10$^{-3}$ . The size of the cluster formed is not uniform, however, the method shows an ability to categorize the patient samples into sub groups that significantly differ in the survival time. In addition, to separating the patient according to survival time with significant statistics the subgroups are correlated to tumor grade.

In addition to mean survival time the clusters also show resembles in two other clinical features that are $vital status$ and $neoplasm cancer status$. The $vital status$ status is categorized into two based on whether the patient current state is deceased or living. Similarly, the $neoplasm cancer status$ is grouped into two patient samples with tumor or tumor free. Note that the missing field in $neoplasm cancer status$ is indicating the unavailability of information.

\begin{table*}
\centering
\caption{Patient stratification using molecular data}
\label{Tab:02}
\begin{tabular}{llllllll}
\toprule
    \multirow{2}{*}{\textbf{Cluster}} &
     \multirow{2}{*}{\textbf{Size}} &
\multirow{2}{*}{\textbf{Avg. age}} &
      \multicolumn{2}{c}{{\textbf{Vital status}}} &
      \multicolumn{3}{c}{{\textbf{neoplasm cancer status}}} \\
         &&& \textbf{Deceased}&\textbf{Living}&	\textbf{With tumor}&\textbf{Tumor free}&\textbf{Missing} \\
\midrule
1&33&61.24&13 (39.39)&20 (60.61)&18 (54.54) &11 (33.33)&4 (12.12)\\
2&48&56.96&27 (56.25)&21 (43.75)&27 (56.25) &10 (20.83) &11 (22.92)\\
3&50&60.34	&31 (62.0)&19 (38.0) &37 (74.0) &10 (20.0) &3 (6.0)\\
4&47&62.47&34 (72.34) &13 (27.66) &36 (76.59) &6 (12.77) &5 (10.64)\\
5&53&59.87&31 (58.49)&22 (41.51)&30 (56.60)&17 (32.08)&6 (11.32)\\
\bottomrule
\end{tabular}
\end{table*}

\begin{table*}
\centering
\caption{Patient stratification using clinical data}
\label{Tab:03}
\begin{tabular}{llllllll}
\toprule
    \multirow{2}{*}{\textbf{Cluster}} &
     \multirow{2}{*}{\textbf{Size}} &
\multirow{2}{*}{\textbf{Avg. age}} &
      \multicolumn{2}{c}{{\textbf{Vital status}}} &
      \multicolumn{3}{c}{{\textbf{neoplasm cancer status}}} \\
         &&& \textbf{Deceased}&\textbf{Living}&	\textbf{With tumor}&\textbf{Tumor free}&\textbf{Missing} \\
\midrule
1&50&49.28&30 (60.0)&20 (40.0)&30 (60.0)&14 (28.0)&6 (12.0)\\
2&64&71.22&38 (59.37)&26 (40.63)&39 (60.94)&17 (26.56)&8 (12.5)\\
3&45&68.24&27 (60.0)&18 (40.0)&30 (66.67)&9 (20.0)&6 (13.33)\\
4&72&52.61&51 (70.83)&31 (43.05)&49 (68.06)&14 (19.44)&9 (12.5)\\
\bottomrule
\end{tabular}
\end{table*}

\begin{table*}
\centering
\caption{Patient stratification with linear kernel weights using molecular data and clinical data}
\label{Tab:04}
\begin{tabular}{llllllll}
\toprule
    \multirow{2}{*}{\textbf{Cluster}} &
     \multirow{2}{*}{\textbf{Size}} &
\multirow{2}{*}{\textbf{Avg. age}} &
      \multicolumn{2}{c}{{\textbf{Vital status}}} &
      \multicolumn{3}{c}{{\textbf{neoplasm cancer status}}} \\
         &&& \textbf{Deceased}&\textbf{Living}&	\textbf{With tumor}&\textbf{Tumor free}&\textbf{Missing} \\
\midrule
1&46&55.07&0 (0.0)&46 (100)&15 (32.61)&25 (54.35)&6 (13.04)\\
2&30&59.23&0 (0.0)&30 (100)&12 (40.0)&15 (50.0)&3 (10.0)\\
3&108	&60.74&108 (100)&0 (0.0)&91 (84.26)&4 (3.70)&13 (12.04)\\
4&24&63.29&24 (100)&0 (0.0)&22 (91.67)&0 (0.0)&2 (8.33)\\
5&23&64.87&4 (17.39)&19 (82.61)&8 (34.78)&10 (43.48)&5 (21.74)\\
\bottomrule
\end{tabular}
\end{table*}

\begin{table*}
\centering
\caption{Patient stratification with optimized kernel weights using molecular data and clinical data}
\label{Tab:05}
 \begin{tabular}{llllllll}
\toprule
    \multirow{2}{*}{\textbf{Cluster}} &
     \multirow{2}{*}{\textbf{Size}} &
\multirow{2}{*}{\textbf{Avg. age}} &
      \multicolumn{2}{c}{{\textbf{Vital status}}} &
      \multicolumn{3}{c}{{\textbf{Neoplasm  cancer status}}} \\
         &&& \textbf{Deceased}&\textbf{Living}&	\textbf{With tumor}&\textbf{Tumor free}&\textbf{Missing} \\
\midrule
1&15&59.53&1 (6.67)&14 (93.33)&4 (26.67)&10 (66.67)&1 (6.67)\\
2&19&63&19 (100)&0 (0.0)&17 (89.47) &0 (0.0)&2 (10.53)\\
3&37&67.38&37 (100)&0 (0.0) &33 (89.19) &0 (0.0)&4 (10.81)\\
4&52&56.60&52 (100)&0 (0.0)&41 (78.85)&3 (5.76)&8 (15.38)\\
5&81&58.09&0 (0.0) &81 (100)&29 (35.80)&40 (49.38)&12 (14.81)\\
6&27&61.11&27 (100)&0 (0.0)&24 (88.89)&1 (3.70)&2 (7.41)\\
\bottomrule
\end{tabular}
\end{table*}

It is observed that the combined molecular data and clinical data although are able to carry out a clear distinction between clusters in terms of mean survival time but lack the distinction in terms of $vital status$ and $neoplasm cancer status$ summarize in Table~\ref{Tab:02} and Table~\ref{Tab:03} respectively.  On the other hand, the results obtained when the clinical data is combined with molecular data are different as reported in Table~\ref{Tab:04} and Table~\ref{Tab:05}. Table ~\ref{Tab:04} shows results for the linear combination of kernels. It is observed that a distinct stratification can be obtained for the $vital status$ , but not in the case of $neoplasm cancer status$. Table ~\ref{Tab:05} reports the results with optimized kernel coefficient used to form the kernel matrix. It is observed that the stratification of the patients is more clear for both $vital status$  and $neoplasm cancer status$.

\subsection{Clinical Outcome Prediction using Molecular Data}
The observations from patient stratification section motivate us to consider integrated molecular data, clinical data, and their combinations. The model is evaluated on a set of dichotomized overall prediction using an LS-SVM.  We carry out a prediction for two characteristic feature survival risk and tumor grade. The survival risk is divided into two based on high risk and low risk periods.
The high risk considers cases where the survival time is lower than median survival time, whereas, low risk considers cases where survival time is higher than median. For the selected samples from TCGA data the median survival time is set to 998 days. We also perform prediction on high and low grade tumor using molecular data, clinical data and their combinations. The low grade contains samples with tumor grade of type G1 or G2 whereas, high grade contains samples corresponding to type G3 and G4 \cite{ncilink} .

\begin{figure*}[t]%figure2
\centering
\includegraphics[clip,trim=0cm 0.1cm 0cm 0cm,height=4cm, width=8.0cm]{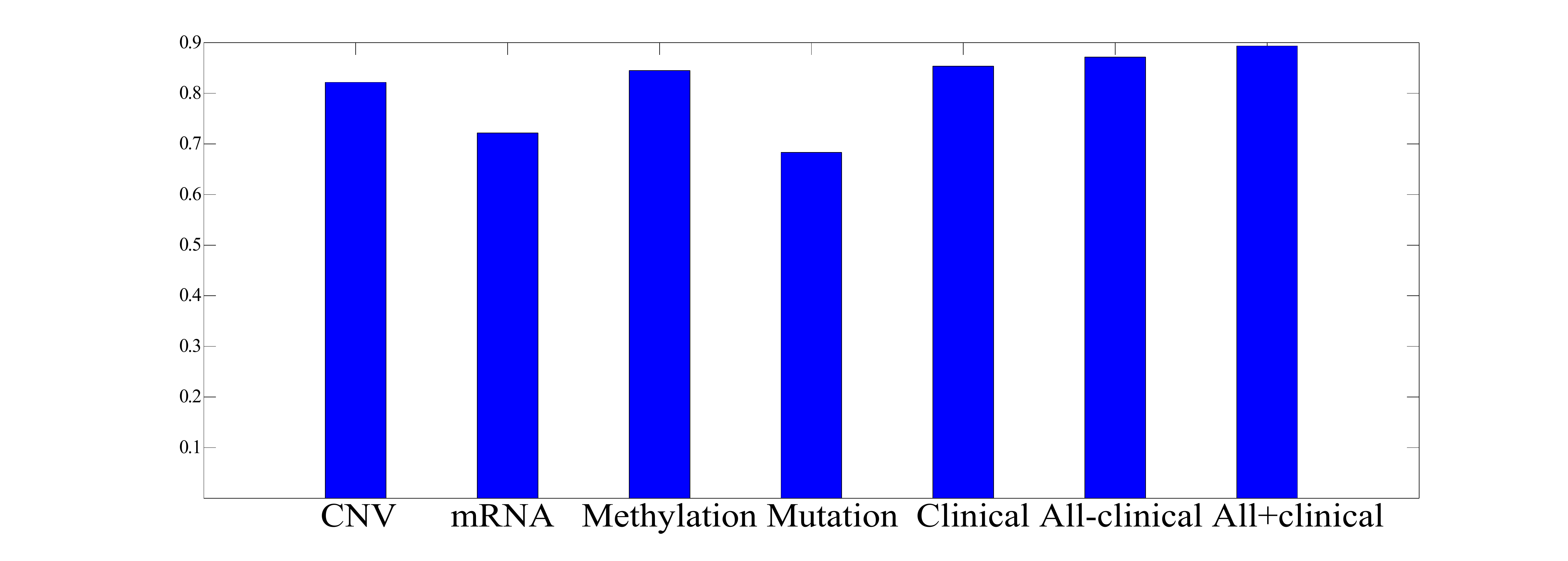}
\caption{ Area under curve for high v.s. low grade classification of OV}
\label{fig:06}
\end{figure*}

Figure~\ref{fig:06} shows the performance behavior of the model summary of the prediction when molecular data are considered in isolation and when all molecular data and clinical features are integrated. It is observed that high AUC values of 0.8217, 0.8449, 0.8538, 0.8718 and 0.8937 are obtained for CNA, methylation, clinical, non-weighted integrated combination and weighted integrated data respectively. These observations also help us to infer some biological information. For the patient samples, in which the changes in tumor samples are due to the structural variation in the chromosome like a copy number variation or methylation seems to have a slightly higher influence on high or low grade clinical predictions. The tumor samples with functional changes like mutation and mRNA also directly relate and contribute towards the classification of clinical outcome. For the developed model, methylation performed better in comparison to other molecular data. Overall, the integration of the different data types improved the prediction accuracy.

\begin{table*}[t]
\centering
\caption{Patient survival risk prediction with individual data types.}
\label{Tab:06}

\begin{tabular}{llllllll}
\toprule
\textbf{Data types} & \textbf{TP}& \textbf{FP} & \textbf{TN} &\textbf{FN} &\textbf{Spec.} & \textbf{Sens.} & \textbf{Acc.}  \\
\midrule
CNA&0.6904&0.3095&0.7368&0.2632& 74.36&68.29&71.25\\
Mutation&0.6667&0.3333&0.7105&0.2895& 71.79&65.85&68.75\\
Methylation&0.6905&0.3095&0.7105&0.2895&72.5&67.5&70\\
mRNA &0.6428&0.3571&0.6842&0.3158&69.23&63.41&66.25\\
Clinical&0.7143&0.2857&0.73684&0.2632&75.0&70.0&72.5\\
\shortstack{CNA+Mutation+mRNA \\Methylation}&0.7381&0.2619&0.7368&0.2632&75.61&71.79&73.75\\
\bottomrule
\end{tabular}
\end{table*}

\begin{table*}[t]
\centering
\caption{Pateint survival risk prediction with individual data types and clinical.}
\label{Tab:07}
\begin{tabular}{llllllll}
\toprule
\textbf{Data types} & \textbf{TP}& \textbf{FP} & \textbf{TN} &\textbf{FN} &\textbf{Spec.} & \textbf{Sens.} & \textbf{Acc.} \\
\midrule
CNA+clinical&0.7143&0.2857&0.7368&	0.2632&75&70&72.5\\
Mutation+clinical&0.6667&0.3333&0.7368&0.2632&73.68&66.67&70\\
Methylation+clinical&0.7143&0.2857&0.7105&0.2895&73.17& 69.23&71.25\\
mRNA+clinical&0.6429&0.3571&	0.7368&0.2632&72.97& 65.12&68.75\\
\shortstack{CNA+Mutation+Methylation \\+mRNA+Clinical}&0.7619&0.2381&0.7632&0.2368&78.05&74.36&76.25\\
\bottomrule
\end{tabular}
\end{table*}

The results for high risk and low risk survival are summarized in Table~\ref{Tab:06} and Table~\ref{Tab:07} with accuracy as the performance measure. The notation used in these tables are TP stands for a true positive, FP stands for a false positive, TN for a true negative, FN for a false negative, Spec. for specificity and Sens. for sensitivity. The accuracy is the proportion of true results (both true positives and true negatives) among the total number of cases examined. Table~\ref{Tab:06} reports the results for molecular data and their integration. It summarizes the behavior of individual data set in prediction accuracy. Amongst the individual molecular data CNA is able to predict low risk and high risk patient for nearly 71$\%$ of the samples. Next, in order to determine the behavior of data type when integrated with clinical data experimentation were carried out, the results are reported in Table~\ref{Tab:07}. The results show an overall increase in accuracy by integration: for the low risk vs. high risk survival classification. These findings are useful as they suggest that some biological information may be fused to various data sources from different genomic levels. Thus, integration of these independent data types increases the chances of success in cancer outcome predictions.

\section{Conclusion}

In this paper,  we have developed a multiple kernel learning based pipeline for integrative analysis of heterogenous data types and apply it on ovarian cancer data. The data types we examined are molecular data and clinical data. The model is used to carry out patient stratification and clinical outcome prediction. We use kernel k-means to perform stratification of patients and examine inter-cluster dissimilarity of survival time, $vital status$  and $neoplasm cancer status$. Stratification is done considering different test cases including integrated molecular data, clinical data, integration using linear non-weighted combination and integration using weighted kernel coefficient combination. The patient stratification results for different test cases show that the integration of molecular and clinical data results in a better pattern forming relation. The clinical outcome prediction is done for tumor grade and survival risk. In the case of tumor grade, a better AUC of 0.8937 was achieved for weighted kernel combination in comparison to 0.8538 considering only clinical data. For survival risk prediction it was observed that when molecular data are integrated with clinical data the overall prediction of the system is improved. This work concludes that integration of molecular data along with clinical data not only helps in carrying out better patient stratification but also improves the prediction accuracy of the model.

%\section*{Acknowledgement}
%This research was supported by Basic Science Research Program through the National Research Foundation of Korea(NRF) funded by the Ministry of Science, ICT and Future Planning (NRF-2015R1C1A2A01055739) and by the KEIT (Korea Institute for Industrial Economics and Trade), Korea, under the "Global Advanced Technology Center" (10053204)

%%%%%%%%%%%%%%%%%%%%%%%%%%%%%%%%%%%%%%%%%%%%%%%%%%%%%%%%%%%%%%%%%%%%%%%%%%%%%%%%%%%%%

%\bibliography{ijmso}
\bibliographystyle{plain}
\bibliography{HetBioDataFusion}

%\end{thebibliography}
\end{document}